# Phase transition in bulk single crystals and thin films of VO$_2$ by nano-infrared spectroscopy and imaging

Mengkun Liu[1,2]*, Aaron J. Sternbach[1], Martin Wagner[1], Tetiana V. Slusar[3], Tai Kong[4], Sergey L. Bud'ko[4], Salinporn Kittiwatanakul[5], M. M. Qazilbash[6], Alexander McLeod[1], Zhe Fei[1], Elsa Abreu[7, 8], Jingdi Zhang[1,7], Michael Goldflam[1], Siyuan Dai[1], Guang-Xin Ni[1], Jiwei Lu[5], Hans A. Bechtel[9], Michael C. Martin[9], Markus B. Raschke[10], Richard D. Averitt[1], Stuart A. Wolf[5, 11], Hyun-Tak Kim[3, 12], Paul C. Canfield[4], D. N. Basov[1]*

[1] Department of Physics, The University of California at San Diego, La Jolla, California 92093, USA.

[2] Department of Physics, Stony Brook University, Stony Brook, New York 11794, USA

[3] Metal-Insulator Transition Creative Research Center, ETRI, Daejeon 305-350, South Korea

[4] Ames Laboratory and Department of Physics and Astronomy, Iowa State University, Ames Iowa, 50010, USA.

[5] Department of Materials Science and Engineering, University of Virginia, Charlottesville, Virginia 22904, USA.

[6] Department of Physics, College of William & Mary, Williamsburg, Virginia 23187, USA

[7] Department of Physics, Boston University, Boston, Massachusetts 02215, USA.

[8] Institute of Quantum Electronics, ETH Zurich, 8093 Zurich, Switzerland

[9] Advanced Light Source Division, Lawrence Berkeley National Laboratory, Berkeley, California 94720, USA.

[10] Department of Physics, Department of Chemistry, and JILA, University of Colorado, Boulder, Colorado 80309, USA

[11] Department of Physics, University of Virginia, Charlottesville, Virginia 22904, USA.

[12] School of Advanced Device Technology, Korean University of Science and Technology, Daejeon 305-333, South Korea

* mengkun.liu@stonybrook.edu
* dbasov@physics.ucsd.edu

# Abstract

We have systematically studied a variety of vanadium dioxide ($VO_2$) crystalline forms, including bulk single crystals and oriented thin films, using infrared (IR) near-field spectroscopic imaging techniques. By measuring the IR spectroscopic responses of electrons and phonons in $VO_2$ with sub-grain-size spatial resolution (~20 nm), we show that epitaxial strain in $VO_2$ thin films not only triggers spontaneous local phase separations but also leads to intermediate electronic and lattice states that are intrinsically different from those found in bulk. Generalized rules of strain and symmetry dependent mesoscopic phase inhomogeneity are also discussed. These results set the stage for a comprehensive understanding of complex energy landscapes that may not be readily determined by macroscopic approaches.

# I. INTRODUCTION

Transition metal oxides (TMOs) host an array of crystal forms among which the material properties can vary dramatically. For example, the local energy landscape of electronic structures can vary from single crystals to poly-domain films [1,2] or strain engineered heterostructures [3]. In addition, the nontrivial charge and orbital orderings in TMOs are highly susceptible to local and external stimuli such as thermal excitation [4], strain [5–7] or light illumination [8–10]. The resultant phase transitions are often accompanied by inevitable mesoscopic electronic and/or magnetic inhomogeneities [11–14], making it extremely difficult to distinguish emergent single phase properties from area-averaging phenomena originating from a mixed state [13,15–17]. As a result, experimental observations strongly depend on the particularities of the samples and measurement techniques, posing difficulties in interpreting the results obtained with complementary experimental approaches. Therefore, a systematic investigation of one model TMO system within distinct crystalline states using one unique but versatile microscopic technique is imperative, and can serve as a solid foothold for future research. Here we utilize infrared near-field methods to disentangle the *local* electron and lattice degrees of freedom in vanadium dioxide.

Vanadium dioxide ($VO_2$) is a canonical correlated electron TMO that has an insulator to metal transition with several orders of magnitude conductivity change accompanied by a first order structural phase transition (monoclinic to rutile) at around 340 K [4]. Under almost six decades of intense studies, $VO_2$ has been fashioned into a diverse range of forms such as bulk crystals [4,18,19], amorphous or highly oriented films [20–22], and nano-/micro-crystals [23–27]. Although the growth of low strain, free-standing single crystals is readily accomplished via solution growth [28,29], the growth of high quality crystalline films has not been possible until recently with developments in epitaxial techniques [21,30]. Amongst these samples, the macroscopically measured metal-insulator transition (MIT) temperature, $T_{MIT}$, can vary by up to 100 K [20,22,29,31,32]. Their electronic and lattice properties also show distinct temperature dependences and anisotropic behaviors [20,26,33–37].

In this work, we are able to compare different crystal forms of this canonical phase transition material under identical test environments using systematic infrared near-field studies. The samples we investigated are solution grown, free-standing, $VO_2$ single crystals and highly oriented epitaxial $VO_2$ films on $TiO_2$ substrates. We show that the global properties of these $VO_2$ samples arise from *a concerted interplay between self-organized phase inhomogeneity and strain modified lattice and orbital reshuffling*.

## II. SCANNING NEAR-FIELD INFRARED MICROSCOPY AND SPECTROSCOPY

Scattering-type scanning near-field optical microscopy (s-SNOM) and Fourier transform IR nano-spectroscopy (nano-IR) [38–41] are capable of performing IR nano-imaging as well as IR broadband nano-spectroscopy. The surface topography, electronic and phonon responses can be obtained simultaneously, a significant advantage over other types of nano-probe techniques. It has been demonstrated as an extremely versatile tool for investigations of many different states of matter—from functional polymers [41,42], to two-dimensional materials [43] and meteorite composites [44]. Furthermore, its unique tip-sample interactions allow one to access and manipulate the photon momentum in addition to the photon energy [45], coupling new degrees of freedom to experiments with flexible probing schemes, as exemplified by recent investigations of thin crystals and interfacial effects [46,47].

Atomic force microscope (AFM) based s-SNOM nano-imaging and broadband nano-IR spectroscopy have an optical spatial resolution limited, to first order, only by the AFM tip radius (~20 nm). The key principle that enables these techniques is the demodulation of the tip scattered signal at higher harmonics ($S_n$, $n \geq 2$) of the tip tapping frequency. In general, a demodulated signal at the second harmonic ($S_2$) is sufficient to achieve ~20nm resolution [48], while the near-field amplitude of higher harmonics such as $S_3$ and $S_4$ can register even better light confinement [49]. Therefore, this procedure effectively discriminates the strongly localized near-field response from background far-field scattering with superior spatial resolution [38–40].

In this work we perform nano-imaging and nano-IR spectroscopy with different light sources to monitor the electronic phase transition and the lattice structural phase transition independently, at representative spectral ranges. For mapping the electronic phase transition, we utilize a table-top quantum cascade laser (QCL) with frequency centered at ~1000 cm$^{-1}$ enabling monochromatic IR imaging (as shown in Fig. 4 - Fig. 6 and Fig. 8 below). Since VO$_2$ reveals an energy gap of about 4000 cm$^{-1}$, IR imaging at 1000 cm$^{-1}$ effectively probes the evolution of the Drude conductivity tail associated with the correlated electrons [14] (see also far-field optical conductivity curves in Supplemental Material [50]). On the other hand, in order to register the phonon responses and hence effectively probe the local structural phase transition, we utilize synchrotron IR light from Beamline 5.4 of the Advanced Light Source (ALS) at Lawrence Berkeley National Lab to perform broadband infrared near-field spectroscopy (broadband nano-IR) (Fig. 7). The high brightness and broadband nature of the synchrotron source enables nano-IR spectroscopy that ranges from 450 cm$^{-1}$ to 4500 cm$^{-1}$, limited only by the detector and beamsplitter combination [51]. This spectral range allows us to address strain reconfigured monoclinic phonon resonances at the nano-scale, probing the essential V-O bond deformation at ~500-750 cm$^{-1}$ with reduced mixed phase ambiguity [52]. (The highest IR active phonon frequency in VO$_2$ reported by experiments and theory is ~750 cm$^{-1}$) [26,50].

We attenuate the output power of the QCL to ~5 mW, which yields a good enough signal-to-noise ratio for S$_3$ and S$_4$ based imaging (third and fourth harmonics of the modulated signal). For nano-IR with the broadband synchrotron source the integrated infrared power (700 – 5000 cm$^{-1}$) is ~0.5 mW, so that we obtain all the broadband nano-IR spectrum with second harmonic demodulation (S$_2$) [51].

### III. SAMPLES

In this work, we examine low strain, solution grown bulk VO$_2$ crystals and highly oriented strained VO$_2$ films on [001]$_R$ TiO$_2$ substrates. We also compare this work to our previous results on highly oriented VO$_2$ films on [110]$_R$ and [100]$_R$ TiO$_2$ substrates, and on low strain polycrystalline VO$_2$ films on (r-cut) sapphire substrates. Despite the obvious variations in in-plane symmetry, these bulk crystals or films also possess dramatically different strain environments which lead to different MIT temperatures. In general compressive (tensile) strain along c$_R$ yields T$_{MIT}$ lower (higher) than in bulk [22,35,36,53,54]. Bulk crystals (undoped) and poly-crystalline films on sapphire substrates usually have a MIT close to 340 K. Films on [001]$_R$ TiO$_2$ substrate have a compressive strain along c$_R$, leading to a T$_{MIT}$ < 340 K [22,54]. Films on [110]$_R$ and [100]$_R$ TiO$_2$ substrates have a T$_{MIT}$ > 340 K with local variations caused by strain inhomogeneities [35,36,55,56]. *It is important to point out that here T$_{MIT}$ represents the global instead of the local M-I transition temperature, as discussed in our previous work* [55]. Different schemes have been suggested to explain the strain induced T$_{MIT}$ variation, including strong electronic correlations [57,58] and orbital control [35,54] but the microscopic mechanism behind the strain modulation remains unclear. Regarding the MIT in general, it is likely that both correlations and structural effects must be taken into account [10,59–61]. Independent of the precise microscopic mechanism, mesoscopic physics certainly is extremely important in these samples. As we will discuss in detail later, strain and symmetry are the two most important parameters for the mesoscopic pattern observed in VO$_2$. We first show the methods of sample fabrication below.

Bulk single crystals of VO$_2$ were grown out of a V$_2$O$_5$-rich binary melt using solution growth techniques [29,62]. Given the relatively low melting point of V$_2$O$_5$, combined with its low vapor pressure and the large, exposed, liquidus line for the formation of VO$_2$, VO$_2$ can be grown out of a solution of excess V$_2$O$_5$. For this work, roughly 8 grams of V$_2$O$_5$ were placed in an amorphous silica tube with 1 gram of lump VO$_2$ that had been synthesized by heating V$_2$O$_5$ under flowing N$_2$ gas. The V$_2$O$_5$ and VO$_2$ were carefully pumped and flushed with Ar gas with intermediate, gentle heating to encourage outgassing of any adsorbed gas. The silica tube was then sealed and heated to 1050 ˚C over 6 hours, dwelled at 1050 ˚C for 3 hours, and then slowly cooled to 775 ˚C over 50 or more hours. At this point the excess V$_2$O$_5$ was decanted from the rod-like VO$_2$ crystals by use of a centrifuge [29]. The single crystalline rods grow along the high temperature, rutile, c-axis and have lengths of up to greater than 1 cm but have cross sectional areas of less than 1 mm$^2$ (in some cases significantly smaller). Selected crystals are shown in the inset of Fig. 1. It should be noted that these crystals have experienced the minimum possible strain given that they grew out of liquid and were removed from this liquid in the high temperature state. When they cooled through the MIT as they approached room temperature they were free-standing in the growth ampule. The signature of the metal to insulator transition in these bulk single crystals can be seen clearly in bulk DC magnetic susceptibility $M(T)/H$ data shown in Fig. 1. The phase transition is very sharp and manifests relatively little hysteresis for heating and cooling rates of roughly 0.2 K per minute. We avoid resistivity measurements here since micro-cracks occurring during heating and cooling through the first order M-I transition are a common drawback for transport measurements in bulk $VO_2$ crystals. When performing the near-field measurements, one end of the crystal is attached to the heating stage with a point contact, thereby minimizing sample holder induced strain.

The highly oriented $VO_2$ films on $[001]_R$ $TiO_2$ substrate ($c_R$ axis out-of-plane, see definition in Fig. 3) were fabricated by the pulsed-laser deposition method. A KrF excimer laser with a radiation wavelength of 248 nm was used to ablate the metallic vanadium target in an oxygen atmosphere. During thin film deposition, the oxygen partial pressure was maintained at 14 mTorr and substrate temperature at 450 ˚C. The deposition rate of the films was about 0.5 nm/min. By adjusting the growth time, we have obtained films with different thicknesses. The temperature dependent resistance of the ~30 nm $VO_2/TiO_2$ $[001]_R$ films are shown in Fig. 2. The maxima in $\left|\frac{1}{\rho}\frac{d\rho}{dT}\right|$ are broad and show substantial hysteresis. Based on this we can set a range for $T_{MIT} \approx 290 \pm 10$ K, with the understanding that this value suggests local, mesoscopic complexities, and is not a "bulk value", as will be discussed later. X-ray characterization (inset of Fig. 2) is also performed, demonstrating the quality (i.e. highly oriented domains) of these films. $VO_2$ films on $[110]_R$ and $[100]_R$ $TiO_2$ substrates were grown by reactive-biased target ion beam deposition [21] and have been previously studied and reported [35,36,55,56,63]. We compared all available films on $TiO_2$ substrates with the same orientations made from different deposition methods and obtained consistent results.

From the various samples investigated in this and previous studies [14,35,36,55,56] we conclude the following for $VO_2$. 1) The MIT in $VO_2$ bulk crystals exhibits a sharp and well-defined phase boundary with $T_{MIT} \approx 341 \pm 2$ K. The hysteresis is narrow due to the minimum strain environment. 2) The MIT in VO$_2$/TiO$_2$ films with out-of-plane $c_R$ axis ([001]$_R$) is dominated by substrate induced lattice clamping, revealing new structural and electronic ordering in a rather uniform manner. The substrate induced compressive strain along $c_R$ yields an isotropic $T_{MIT} < 340$ K and a wide hysteresis. 3) The MIT in VO$_2$/TiO$_2$ films with in-plane $c_R$ axis ([100]$_R$ or [110]$_R$)) is dominated by mesoscopic, anisotropic phase separation. Substrate induced tensile strain along $c_R$ leads to macroscopic electron transport anisotropy with a $T_{MIT} > 340$ K when measured along $c_R$. ($T_{MIT} < 340$K perpendicular to $c_R$). The thicknesses of all the thin films we investigated were greater than 30nm (range from 30nm to 300nm). Therefore, we cannot exclude the possibility that for much thinner films (e.g. a few nm) one could observe different microscopic patterns and strain induced behaviors. For clarity we illustrate the lattice and unit cell structures of monoclinic M1, monoclinic M2 and rutile R phases in Fig. 3 [57,59,64,65]. In the M1 phase all V atoms pair and twist to form V-V dimers, while in the M2 phase half of the V atoms pair with the other half remaining unpaired [59]. The corresponding lattice axes are also shown in Fig. 3 as well. We note that our current characterization and categorization of the mesoscopic patterns in VO$_2$ are not exhaustive due to the diversity of crystal forms and to the current lower limit in IR frequency one can easily detect at the nano-scale (>500cm$^{-1}$).

## IV. INFRARED NANO-IMAGING OF VO$_2$ SINGLE CRYSTALS AND VO$_2$/[001]$_R$ TiO$_2$ FILMS

Figure 4 illustrates the distinct differences between bulk VO$_2$ crystals and highly oriented VO$_2$ film behavior, investigated by nano-IR imaging at ~1000 cm$^{-1}$. For bulk crystals, there is an abrupt phase front that separates the different phases (Figs. 4 (a)-(f)). From room temperature (RT) to ~334 K the entire sample is in the M1 monoclinic insulating state (Fig. 4 (a)). Above 334 K, a striped monoclinic insulating M2 phase emerges and propagates from the bottom to the top of the field of view as the temperature increases, until at ~336K the entire sample (~30 μm × 100 μm × 3 mm) becomes M2 (Figs. 4 (b) and (c)). At 339.96 (±0.03) K the metallic state emerges within the field of view of our nano-IR setup and it propagates relatively fast with very small temperature increments (Figs. 4 (d) and (e)). At 340.10 (±0.03) K the crystal becomes metallic within our field of view. At each temperature step we wait ten minutes after the sample is fully equilibrated prior to the measurement. Figure 4 (a) - (f) clearly demonstrate that bulk crystals have well-defined phase boundaries that separate different states (see also Fig. S2 with enhanced contrast in the Supplemental Material [50]). The angle of the insulator-metal phase boundary is approximately 50 degrees with respect to the $c_R$ axis, and can be affected by local impurities on the surface. We obtained essentially the same domain orientation at larger scales with an optical microscope as those reported in references [66,67] where the bulk $VO_2$ samples were thermally excited by an electric current. Therefore, the results of large scale characterization of the single crystal samples will not be covered in detail here. Our assignment of the M1 and M2 phase is confirmed by Raman spectroscopy (not shown), and further investigated with the broadband nano-IR studies shown in Fig. 7.

Figures 4 (g)-(l), on the other hand, show simultaneous near-field and AFM images of a 30nm $VO_2/[001]_R$ $TiO_2$ film at room temperature. The near-field images are again measured at ~1000 cm$^{-1}$. Due to an intrinsic topographic buckling (e.g. Fig. 4 (h)), the films we used in this study are shaped into "microbeams" with different lateral widths, leading to various strain environments. In the vicinity of the buckling (or sometimes cracking) the strain in the $VO_2$ film is effectively released, we found that for narrower beams the compressive epitaxial strain (along $[001]_R$) is overall reduced, leading to a higher local transition temperature, closer to 340 K. This is clearly identified in Fig. 4 (g): there are multiple scattered near-field signal levels reflecting step-like variations of IR conductivities between adjacent beams of different widths. The AFM amplitude (phase) in Fig. 4 (k) (Fig. 4 (l)) and the near-field amplitude (phase) in Fig. 4 (i) (Fig. 4 (j)) within the same microbeam are uniform and smooth, showing no mixture of "metallic" and "insulating" states within our 20 nm spatial resolution. It is unlikely that there is a depth-dependent conductivity variation since the film thickness is only ~30 nm. This indicates that the variety of near-field signals observed in our $VO_2$ films (Fig. 4(g)) most likely originate from distinct stages of the MIT being stabilized in different domains of the film for a given temperature. We note that the microbeam width dependent strain environment has been reported for $VO_2/TiO_2$ $[100]_R$ films [55], in which the substrate provides an alternating local strain (along $c_R$) environment instead of an overall compressive strain mismatch, as is the case for $VO_2/TiO_2$ $[001]_R$ films.

To further illustrate the differences between bulk crystals and films on $[001]_R$ $TiO_2$, we compare line profiles and histogram statistics of near-field IR signals in Fig. 5 and Fig. 6, respectively [68]. Line profiles of AFM topography and near-field amplitude $S_3$ of $VO_2$ bulk crystals reveal a sharp electronic as well as structural phase boundary with a width of the order of 100 nm (Figs. 5 (a) - (c)). The insulating M2 state is observed with a reversible stripe-like deformation in topography, in agreement with previous reports [53,69,70]. Line profiles of $S_3$ and topography in 100 nm thick $VO_2/[001]_R$ $TiO_2$ films, however, reveal a terrace-like or gradual change of $VO_2$ metallicity (Figs. 5 (d) - (f)). Multiple near-field $S_3$ signal levels are clearly evident with a dependence on the strain environment, defined by the distance between buckles (1→6 in Fig. 5 (f)), in agreement with Fig. 4 (g). Notice that at certain edges of the $VO_2$ beams the scattering signal can change continuously, indicating a gradual strain relief towards the edges or corners of the beams (6→5). It is worth emphasizing the differences between the topography in bulk crystals and in $VO_2/TiO_2[001]_R$ thin films. In bulk crystals the striped M2 phase (Figs. 3 (a) and (b), and Figs. 5 (a) and (b)) is reversible with temperature (the stripe will disappear with the M2 → M1 transition) and is spontaneously formed at ~335K. In contrast, the buckles in the particular $VO_2$ thin films (Figs. 4 (g) - (i), and Figs. 5 (d) and (e)) we investigated were formed during the film deposition and are therefore permanent.

The histogram of the near-field amplitude $S_3$ gives a straightforward statistical representation of the metallicity at different stages of the phase transition for both samples (Fig. 6). Bulk single crystals present binary statistics (in our case, dark blue or bright yellow in near-field contrast) everywhere on the samples, revealing that the samples are in either the fully insulating or the fully metallic state (Figs. 6 (a)-(c)). However, the strained films can form a continuous gradient of scattering signals over the sample. In Figs. 6 (d)-(e) we measured at two adjacent locations close to a strain relieved edge (to the left of Fig. 6 (d), not shown) in the 100 nm $VO_2/TiO_2$ $[001]_R$ film. The continuous histograms in this case clearly reveal an evolution of the intermediate metallic state that advances toward a fully metallic state under high epitaxial strain.

## V. BROADBAND NANO-IR OF VO$_2$ SINGLE CRYSTALS AND VO$_2$/[001]$_R$ TiO$_2$ FILMS

To carefully map out the subtle changes in lattice and electronic states among our samples at the nano- and mesoscopic scales, we rely on the spectroscopic signatures that are representative of the structural and electronic properties. Broadband nano-IR experiments with a lateral spatial resolution below 20 nm were performed at ALS for the bulk crystals and thin films. The IR active phonon responses of the M1 and M2 phases are taken at 3 adjacent locations across the M1-M2 interface in the single crystal at 337 K (see topography in Fig. 7 (a)), and the spectral amplitude [51] is plotted in Fig. 7 (b). When going through the M1 to M2 phase transition, VO$_2$ is known to acquire a slight tensile strain along the $c_R$ axis [71] (see also the phase diagram in Fig. 8 (a)). Accordingly, we observe a blue shift of the ~520 cm$^{-1}$ phonon peak. The peak positions corresponding to the phonon response of M1 are consistent with previous experimental work reporting A$_u$ phonons at ~521 cm$^{-1}$, ~607 cm$^{-1}$ and ~637 cm$^{-1}$ [26,52]. On the other hand our infrared measurements of a pure M2 phonon has not, to our knowledge, been previously reported. While single crystals are often regarded as classical strain free samples and we have indeed made great efforts to grow them in a strain free environment, the M1 phase (strain-free) and M2 phase (strained) can nevertheless coexist under ambient conditions, generating complexities in terms of the spontaneous phase inhomogeneities (striped M2 state). We expect this self-organized strain environment to be more complicated for highly doped single crystals [29] and crystals under uniaxial stress [57,72] or high hydrodynamic pressure [72,73], since several different monoclinic states (M1, M2, T) can coexist. This could be the reason why there are ambiguous "two-phase" regions observed in early experiments on bulk crystal VO$_2$ crystals [19].

In VO$_2$/TiO$_2$[001]$_R$ films, the spectroscopic transition from M1 insulator to rutile metal is clearly demonstrated as a gradual evolution in both electronic and lattice response (Fig. 7

(d)). The broadband nano-IR scans are performed at the center of the microbeams with different beam widths (at room temperature), yielding different strain environment. The locations of representative probe spots (three out of five) with ~20 nm resolution are indicated in Fig. 7 (c). With increasing compressive strain (wider beam width) along $c_R$ the M1 phonon response at ~540 cm$^{-1}$ ($B_u$ mode) gradually red shifts, as indicated by the red arrow in Fig. 7 (d). At the same time, the metallicity also gradually increases, as indicated by the black arrow at ~950 cm$^{-1}$. We note that we observe another $B_u$ mode at ~750 cm$^{-1}$, whose magnitude decreases with strain but whose frequency shift shows no consistent behavior. In this work we focus on the ~540 cm$^{-1}$ resonance. The strain levels for each the microbeams we investigated with broadband nano-IR are roughly estimated in the VO$_2$ phase diagram of Fig. 8 (a), as will be discussed later. We note that the curves in Fig. 7 (d) are vertically shifted for clarity and that the original positions of each curve are indicated in the ~960 cm$^{-1}$-1000 cm$^{-1}$ range. Since the ~540 cm$^{-1}$ monoclinic phonon reflects the V-O bond length, the prominent red shift is in agreement with previous experiments which suggested that V-O bond lengths are increased under compressive strain along $c_R$ [54]. The increasing metallicity is also in line with the statement that orbital occupancies are dramatically altered under strain [35,54]. The results of Figs. 7(c) and (d) indicate that strain tuning and the corresponding lattice and electronic responses can be gradual, revealing intrinsic pure phase properties and ruling out ambiguous averaging effects within our spatial resolution.

## VI. SUMMARY OF MESOSCOPIC INHOMOGENEITY IN VO$_2$ SINGLE CRYSTALS AND VO$_2$ FILMS ON FOUR DIFFERENT SUBSTRATES

To illustrate the strain environment and corresponding phase transitions in all of the samples, we map each sample onto the phase diagram in Fig. 8 (a). Bulk crystals have the narrowest strain span (blue horizontal arrow) indicated by a sharp MIT around 340K (Fig. 1 and Fig. 2). VO$_2$ films on [001]$_R$ TiO$_2$ have a compressive strain along $c_R$ that can lead to a continuous red shift of the phonon response and increase in the IR conductivity (red arrow). The colored dots correspond to the estimated strain environment at which the broadband nano-IR data are taken in Fig. 7. We would like to emphasize that our spatial resolution is sufficient to report on single-phase properties. That is, each dot on the phase diagram corresponds to a new structural or orbital state and there is no "mixed-phase" signal. For comparison, we also illustrate the strain environment in samples with the $c_R$-axis in-plane, e.g. samples on $[100]_R$ and $[110]_R$ TiO$_2$ substrates, as previously studied (green dashed double arrow) [36,55]. For these samples, an anisotropic stripe state is induced with alternating tensile and compressive strain (see also Fig. 8 (b)) [36,56]. As a result, a combination of phase separation and strain-induced modification of orbital occupancy is expected to play a very important role in these films.

Figure 8 (b) sums up the representative mesoscopic domain textures in some of the VO$_2$ samples we measured. For VO$_2$ bulk crystals (blue panel), a sharp insulator-to-metal phase boundary is observed between the M and R phases. For highly oriented VO$_2$ films on $[110]_R$ TiO$_2$ with in-plane $c_R$-axis (green panels), self-organized patterns are evident. In these films, an alternation of compressive and tensile strain yields interesting stripe-like patterns due to the epitaxial strain mismatch and VO$_2$ symmetry breaking [36,55]. For highly oriented VO$_2$/TiO$_2$ $[001]_R$ films with out-of-plane $c_R$-axis (red panel), we observe a gradual phase transition with temperature, as detailed in this study. No stripe-like pattern is observed. For comparison, we also include previously reported results for VO$_2$ films on r-cut sapphire substrates [74], where the samples can have several preferred growth directions due to large lattice mismatch.

While the near-field images in Fig. 8 (b) reveal the substantial phase complexity of VO$_2$, we have nevertheless been able to identify distinct trends that assist in understanding the mesoscopic inhomogeneity observed with nano-imaging and broadband nano-IR. These trends are summarized in the x- and y- axes in Fig. 8 (b), and described using global in-plane symmetry and local strain parameters. For clarity, we first introduce the notion of global in-plane symmetry (bottom x- axis of Fig. 8 (b)). Global in-plane symmetry includes both the crystal symmetry of the sample and the orientation of the substrate. It changes with different substrates or with different film orientations- even though the film is always composed of the same material. For example, for VO$_2$ films on r-cut sapphire substrates (lower left panel in Fig. 8 (b)), the films have several in-plane $c_R$ orientations (low symmetry) due to lattice symmetry mismatch between the sample and the substrate.

This results in phase percolation without clear mesoscopic patterns (heterogeneous). Highly oriented films with one specific in-plane $c_R$-axis (intermediate symmetry) yield stripe-like (anisotropic) patterns (for example, $VO_2$/$TiO_2$ $[110]_R$ in Fig. 8 (b), top middle panel) [36,56] and can have spontaneous symmetry breaking defined by the crystal axis of the $VO_2$ films [55]. For films with the $c_R$-axis out-of-plane (highest in-plane symmetry) the mesoscopic response is rather uniform (homogenous; for example, $VO_2$/$TiO_2$ $[001]_R$, top right panel). Besides in-plane symmetry, one should also consider local strain (left y-axis in Fig. 8 (b)), which is distinct from hydrostatic pressure applied to the samples. Local strain is related to the global film-substrate symmetry due to corresponding strain mismatch but can also be adjusted independently through, for example, film thickness and strain relief due to buckles, cracks or defects. It changes the intrinsic structural and electronic configurations (e.g. $VO_2$/$TiO_2$ $[001]_R$, top right panel, and Ref [54]) while also resulting in a spatial phase redistribution so as to minimize the local strain energy (e.g. $VO_2$/$TiO_2$ $[110]_R$, top middle panel; notice that strain decrease with increasing film thickness and that the mesoscopic stripe periodicity is therefore scaled accordingly).

## VII. SUMMARY AND OUTLOOK

In conclusion, phase separation and structural and electronic reconfigurations can occur concertedly in vanadium dioxide, controlled by the local strain and global symmetry. Our results are specific to $VO_2$, yet universally relatable to other correlated electron materials [75]. Previously studied $VO_2$ nano-beams and micro-crystals can be regarded as specific cases with some of the unique strain environments identified in the present study [23–25,76]. In future studies, ultrafast femtosecond dynamics [77–79] and magnetic orders [1] can also be included in systematic near-field investigations of TMOs, which will lead to a better understanding of the rich diversity of mesoscopic and microscopic phenomena at fundamental time and length scales in materials such as manganites and high temperature superconductors.

ACKNOWLEDGMENTS

D. N. B. acknowledges support from ARO under Grant No. W911NF-13-1-0210. M. L. and M. G. acknowledge support from Helmholtz Virtual Institute MEMRIOX. Development of nano-optics capabilities at UCSD is supported by DOE-BES under Grant No. DE-SC0012592. R. D. A. and E. A. acknowledge support from DOE-BES under Grant No. DE-FG02-09ER46643. A. S. M. acknowledges support from the Basic Energy Sciences initiative of the U.S. Dept. of Energy (DOE-BES). The Advanced Light Source is supported by the Director, Office of Science, Office of Basic Energy Sciences, of the U.S. Department of Energy under Contract No. DE-AC02-05CH11231. P. C. C acknowledges support from the U.S. Department of Energy, Office of Basic Energy Science, Division of Materials Sciences and Engineering for the research performed at the Ames Laboratory. Ames Laboratory is operated for the U.S. Department of Energy by Iowa State University under Contract No. DE-AC02-07CH11358. H.T.K acknowledges support from the creative project in ETRI. S. K., J. L., and S. A. W. are grateful for the support from the Nanoelectronics Research Initiative (NRI) and VMEC. M. M. Q acknowledges support from NSF DMR (grant # 1255156) and the Jeffress Memorial Trust. We acknowledge Rob Olmon for valuable discussions and thank him for developing synchrotron infrared near-field spectroscopy (SINS) at the ALS.

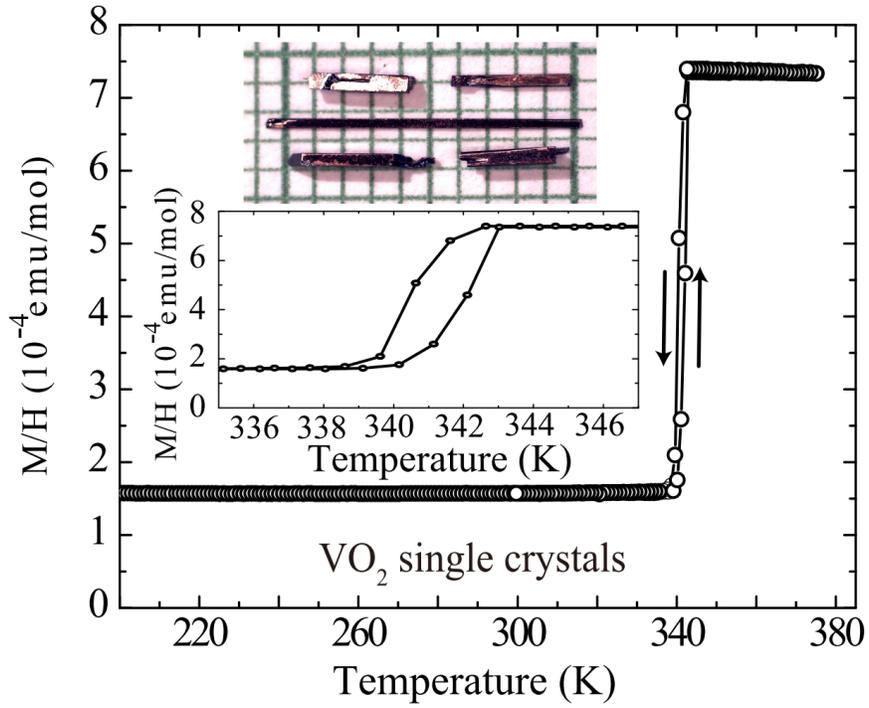

Fig. 1. (color online) (a) Temperature-dependent DC magnetization susceptibility of bulk single crystal $VO_2$. $T_{MIT} \approx 341 \pm 2$ K. Inset, top: microscopic image of selected $VO_2$ single crystals; Bottom: magnetization curve expanded in the vicinity of the phase transition.

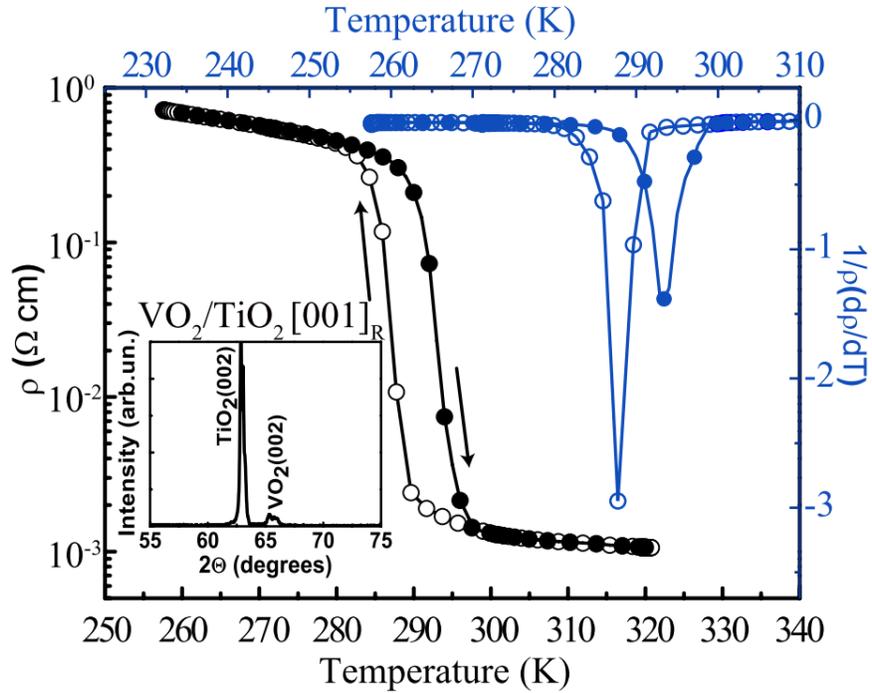

Fig. 2. Temperature dependent resistivity (black) and corresponding derivative (blue) curves for 30 nm $VO_2/TiO_2$ $[001]_R$ film. $T_{MIT} \approx 290\pm10$ K. Inset, x-ray diffraction intensity (arbitrary units). The solid and open circles represent temperature ramping up and down, respectively.

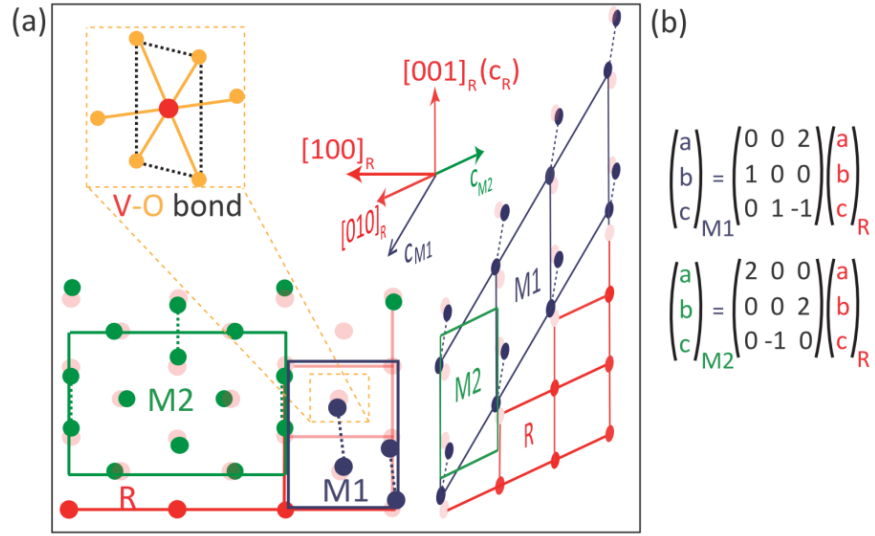

Fig. 3. (color online) (a) Schematic of real space lattice structures of monoclinic M1, M2 and rutile R phases. Blue (green) filled circles: vanadium atoms in M1 (M2) phase; Red filled circles: vanadium atoms in R phase. Blue (green) solid line: unit cell of M1 (M2) phase. Blue (green) dashed line: V-V pairing in M1 (M2) phase. Red solid line: unit cell of R phase. The arrows indicate the corresponding lattice axes. (b) Matrix equation of M1-R, M2-R lattice relationship.

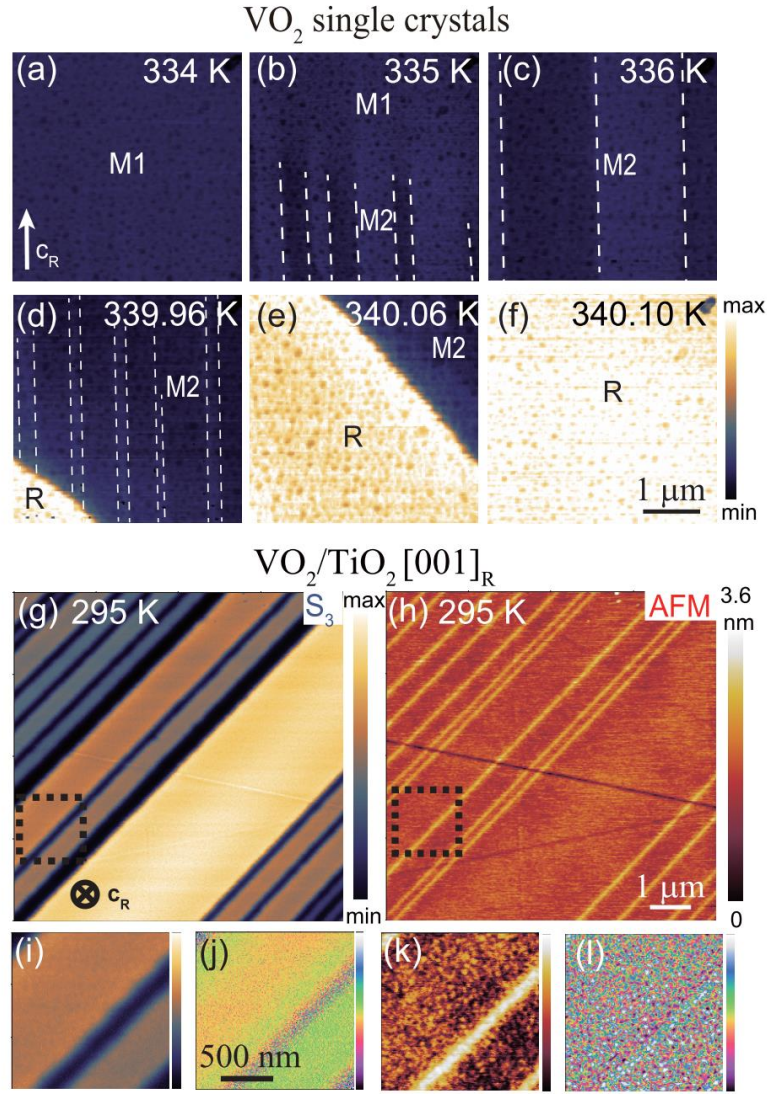

Fig. 4. (color online) (a)-(f) IR scattering amplitude $S_3$ of $VO_2$ bulk single crystals acquired at different temperatures. Note that the temperature difference from (d) to (f) is only ~0.14 K (±0.03K). The stripes in the M2 phase are depicted by the white dashed lines. For enhanced contrast of (a)-(f) see Fig.S2 in supplemental materials [50]. The (g) and (h), simultaneously acquired scattering $S_3$ and AFM images of 30nm $VO_2/TiO_2$ $[001]_R$ film. The dashed squares in (g) and (h) indicate the area whose expanded view is shown in (i)-(l). (i), near-field $S_4$ amplitude; (j) near-field $S_4$ phase; (k) AFM amplitude; (l) AFM phase. The IR probing frequency (from a QCL laser) is 1000 cm$^{-1}$. The measurements for panel (g)-(l) are taken at 295 K.

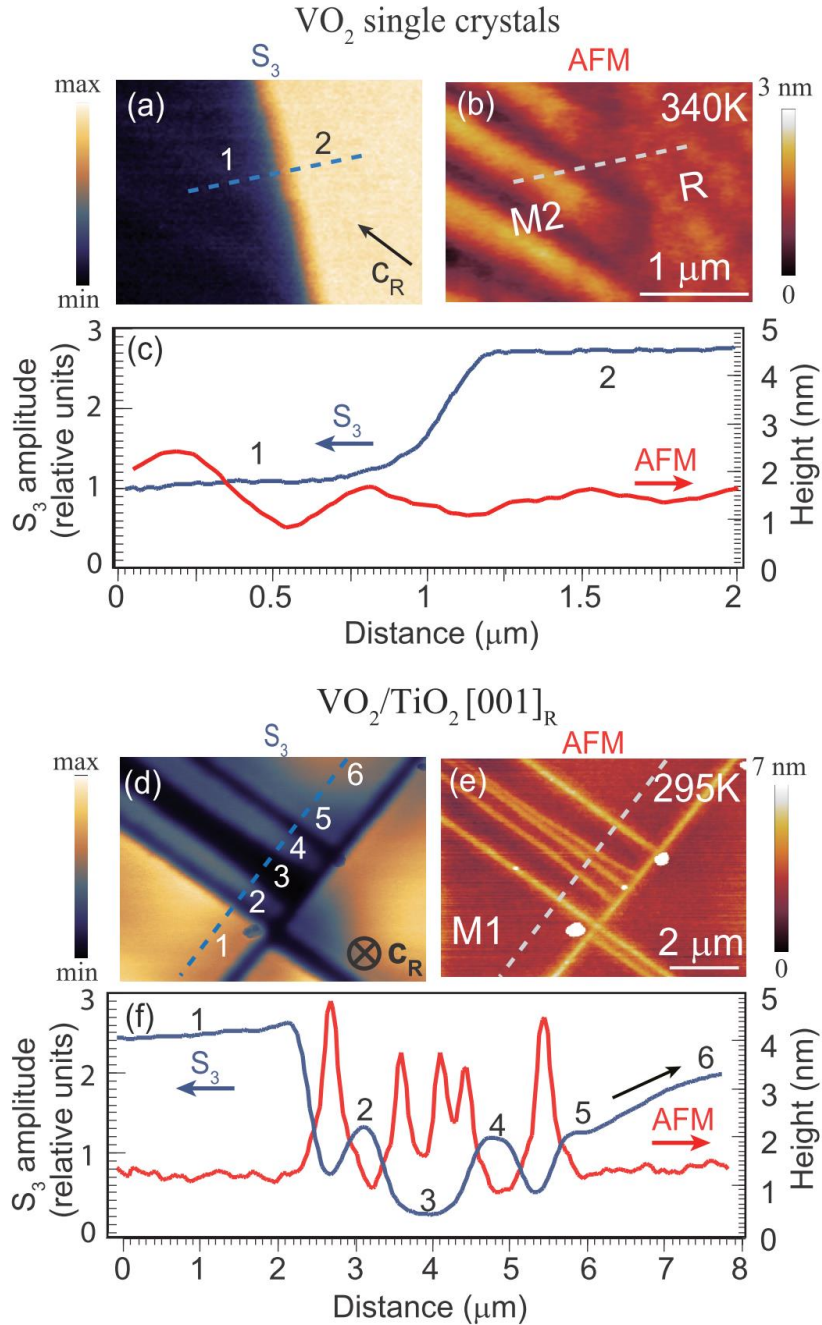

Fig. 5. (color online) Near-field amplitude $S_3$ (a) and AFM (b) images of $VO_2$ bulk crystals at 340K. (c) Line profiles of near-field amplitude $S_3$ (blue) and topography (red) taken along the dashed lines in (a) and (b), respectively. (d) and (e) Near-field amplitude $S_3$ and AFM topography of 100nm $VO_2/TiO_2$ $[001]_R$ film. The substrate induces a strain dependent $VO_2$ metallicity. (f) Line profiles of the near-field amplitude $S_3$ (blue) and topography (red) taken along the dashed line in (d) and (e), respectively.

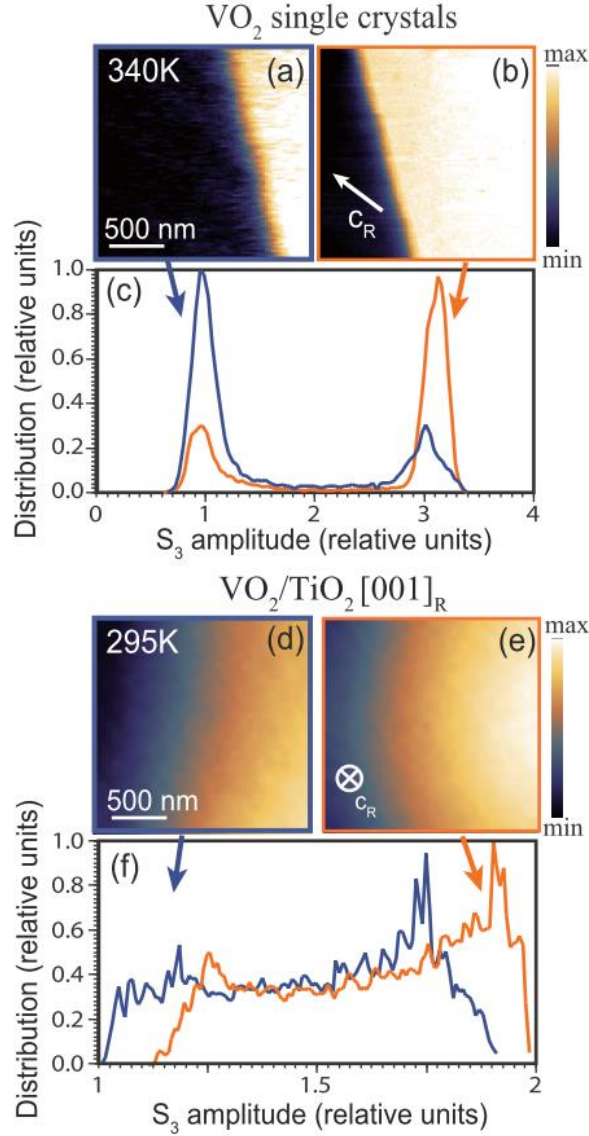

Fig. 6. (color online) IR scattering amplitude ($S_3$) of $VO_2$ single crystals at (a) less evolved and (b) more evolved MIT regions at 340K. (c) Histograms of (a) and (b) show a bimodal insulator to metal phase transition. In contrast, 100nm $VO_2/TiO_2$ $[001]_R$ film at (d) less strained and (e) more strained regions of the film at 295 K exhibit a rather gradual and continuous MIT transition. The strain relieved buckled edge is close to the left edge of Fig. 6 (d) (not shown) (f) Histograms show gradual phase transitions spanning almost uniformly across the images, indicating a continuous evolution of the intermediate insulator-metallic states. All figures have the same scale. Scale bar: 500nm.

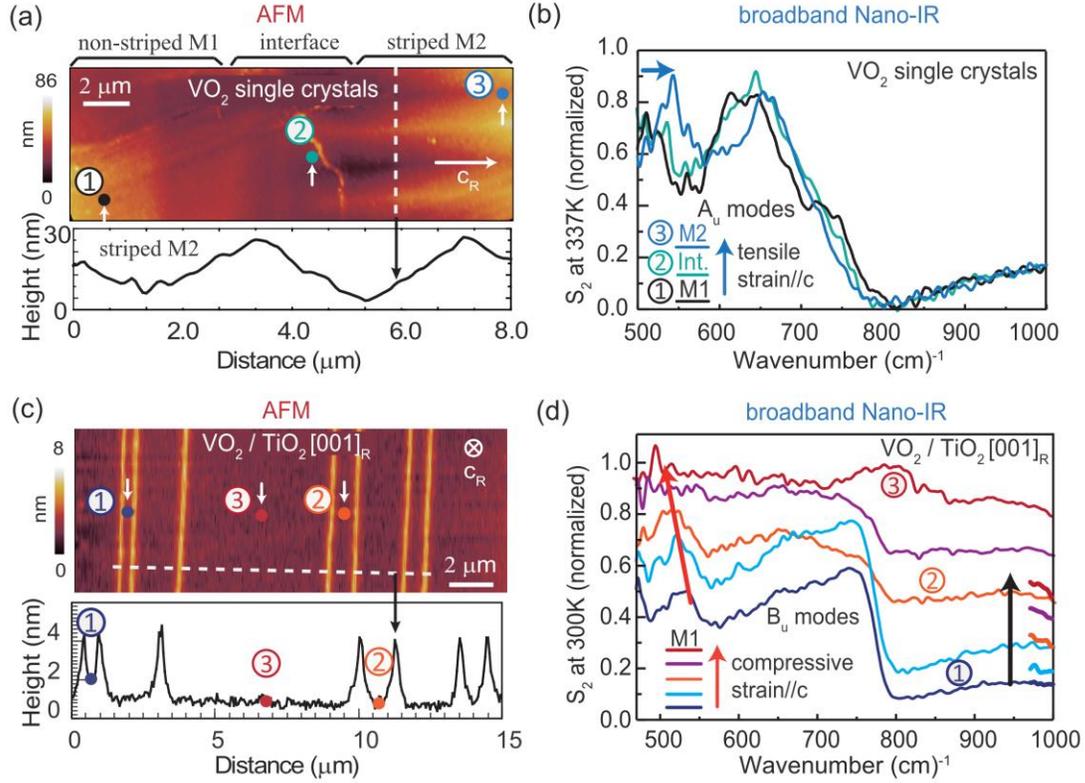

Fig. 7. (color online) (a) top: AFM topography of a M1-M2 interface in VO$_2$ single crystal at 337K; bottom: AFM line scan of the striped M2 phase (along the white dashed line in the top panel). (b) Normalized nano-IR spectra S$_2$ at different locations corresponding to the spots indicated by the short white arrows in Fig. 7(a). Since the c$_R$ axis is in-plane (white horizontal arrow in Fig. 7(a)), the major contribution of the signal is from A$_u$ phonon modes (E//a$_R$). (c) Top: AFM topography of the 100nm VO$_2$/TiO$_2$ [001]$_R$ film at room temperature. Bottom: AFM line scans of the buckles and the microbeams (along white dashed line in the top panel). (d) Normalized nano-IR spectra S$_2$ taken at the center of the micro beams with different beam widths. Three out of the five representative spectra are taken at the probe spots indicated by the short white arrows in (c). Since the c$_R$ axis is out-of-plane, the major contribution of the signal is from B$_u$ phonon modes. For clarity, the curves in (d) are shifted vertically. Original positions of each curve are indicated at the high frequency end of the same panel. We note that since the tip scattered IR wave-vector and its polarization is not well defined, the assignments of A$_u$ and B$_u$ modes in (b) and (d) is qualitative. We also note that all the S$_2$ spectra shown in Fig. 7 (b) and (d) have a ~20 nm spatial resolution.

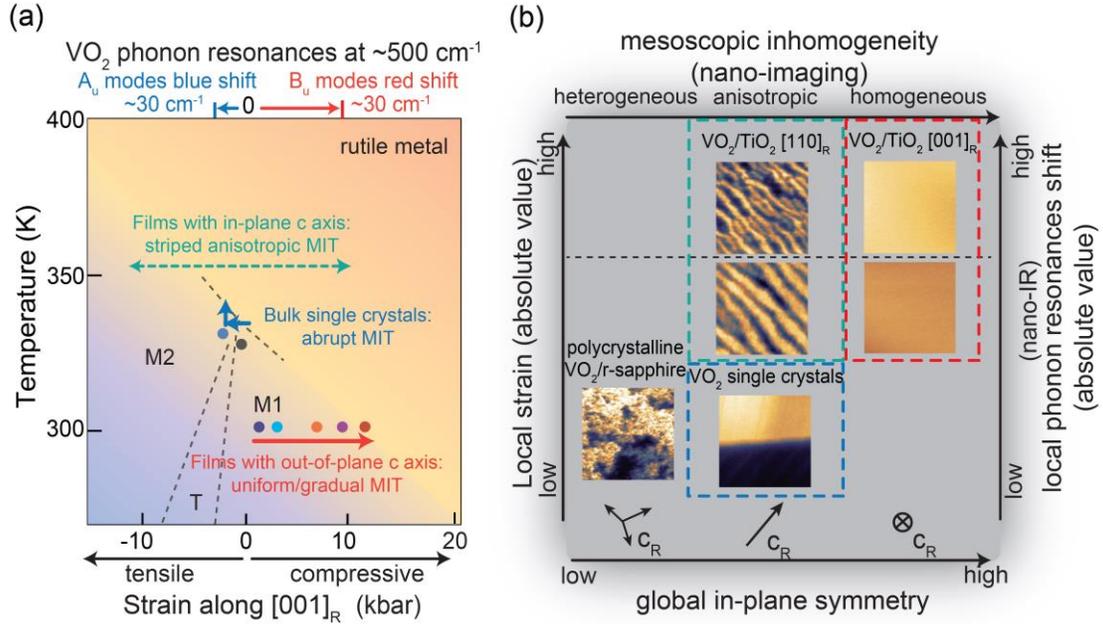

Fig. 8. (color online) (a) VO$_2$ temperature-strain phase diagram. The vertical position of the arrows indicates the global T$_{MIT}$ of corresponding VO$_2$ samples. The horizontal span of the arrows indicates the extent of the strain environment in the samples. The phonon resonance blue shifts in single crystals and red shifts in films with out-of-plane c$_R$ axis, with increasing tensile and compressive strain along c$_R$, respectively, as indicated at the top of Fig. 8 (a). The films are referring to VO$_2$ films on TiO$_2$ substrates. (b) Generalized strain-symmetry-inhomogeneity trends of VO$_2$ crystals and films. With increasing global in-plane symmetry (see text), the mesoscopic patterns probed by nano-imaging become more homogeneous (heterogenous→ anisotropic → homogeneous). With increasing local strain, the periodicity of the stripe-like pattern (in the case of VO$_2$/TiO$_2$ [110]$_R$ [36,55]) or strain induced metallicity (in the case of VO$_2$/TiO$_2$ [001]$_R$) changes accordingly. The phonon resonance shift probed by broadband nano-IR is a good indicator of the local strain which modifies the lattice structure. Bottom left panel: 100nm VO$_2$/r-sapphire at 343K; Top middle: 50nm VO$_2$/TiO$_2$ [110]$_R$ at 340K (higher local strain); middle: 100nm VO$_2$/TiO$_2$ [110]$_R$ at 332K (lower local strain); bottom middle: VO$_2$ single crystal at 340K; top right: 100nm VO$_2$/TiO$_2$ [001]$_R$ at 300K (selected scan window far away from buckles, higher local strain); middle right: 100nm VO$_2$/TiO$_2$ [001]$_R$ at 300K (selected scan window closer to buckles, lower local strain). All images in (b) are 3 μm x 3 μm, taken with near-field amplitude S$_3$ except for VO$_2$ on sapphire (S$_2$).

# Supplemental material for
# Phase transition in bulk single crystals and thin films of VO$_2$ by nano-infrared spectroscopy and imaging

Mengkun Liu[1,2] *, Aaron J. Sternbach[1], Martin Wagner[1], Tetiana V. Slusar[3], Tai Kong[4], Sergey L. Bud'ko[4], Salinporn Kittiwatanakul[5], M. M. Qazilbash[6], Alexander McLeod[1], Zhe Fei[1], Elsa Abreu[7,8], Jingdi Zhang[1,7], Michael Goldflam[1], Siyuan Dai[1], Guang-Xin Ni[1], Jiwei Lu[5], Hans A. Bechtel[9], Michael C. Martin[9], Markus B. Raschke[10], Richard D. Averitt[1], Stuart A. Wolf[5,11], Hyun-Tak Kim[3,12], Paul C. Canfield[4], D. N. Basov[1*]

[1] Department of Physics, The University of California at San Diego, La Jolla, California 92093, USA.

[2] Department of Physics, Stony Brook University, Stony Brook, New York 11794, USA

[3] Metal-Insulator Transition Creative Research Center, ETRI, Daejeon 305-350, South Korea

[4] Ames Laboratory and Department of Physics and Astronomy, Iowa State University, Ames Iowa, 50010, USA.

[5] Department of Materials Science and Engineering, University of Virginia, Charlottesville, Virginia 22904, USA.

[6] Department of Physics, College of William & Mary, Williamsburg, Virginia 23187, USA

[7] Department of Physics, Boston University, Boston, Massachusetts 02215, USA.

[8] Institute of Quantum Electronics, ETH Zurich, 8093 Zurich, Switzerland

[9] Advanced Light Source Division, Lawrence Berkeley National Laboratory, Berkeley, California 94720, USA.

[10] Department of Physics, Department of Chemistry, and JILA, University of Colorado, Boulder, Colorado 80309, USA

[11] Department of Physics, University of Virginia, Charlottesville, Virginia 22904, USA.

[12] School of Advanced Device Technology, Korean University of Science and Technology, Daejeon 305-333, South Korea

*Corresponding author: mengkun.liu@stonybrook.edu (M.K.Liu); dbasov@physics.ucsd.edu (D.N.Basov)

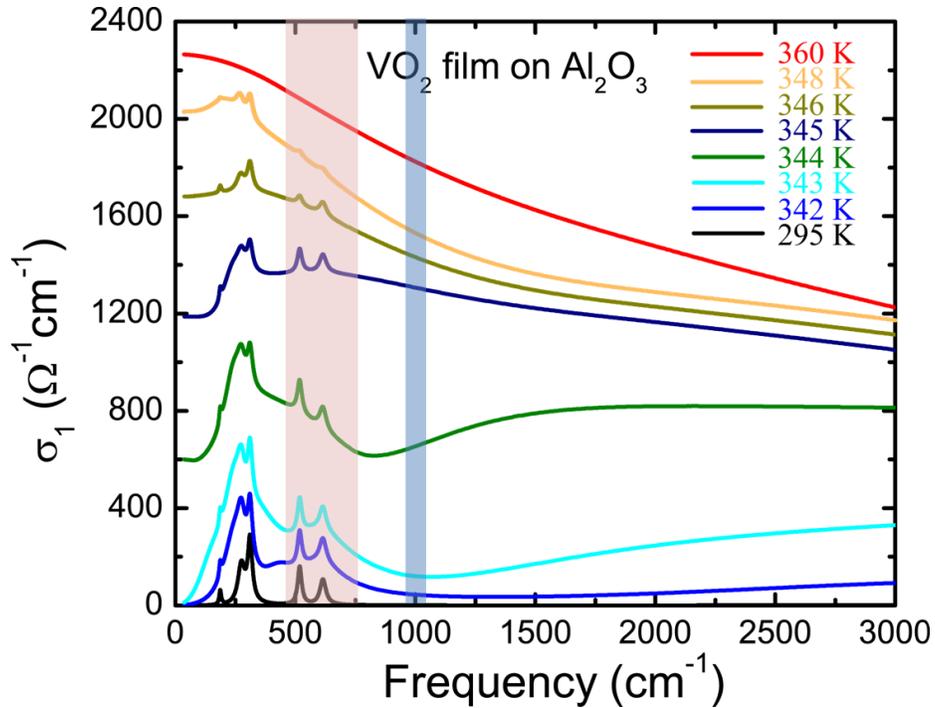

**Figure S1 Fourier transform infrared spectroscopy (far-field spectrum) of VO₂ film on sapphire** [Science 318, 1750 (2007), Phys. Rev. B 79, 075107 (2009)]. *This figure represents a general idea of how we assign our near-field measurements at different frequency ranges.* We perform near-field imaging (Fig. 4 - Fig. 6 and Fig. 8 in the manuscript) at ~1000 cm$^{-1}$ (blue bar) to measure the electronic Drude responses. 1000 cm$^{-1}$ is far enough away from the highest phonon response (~750 cm$^{-1}$) and well below the insulating bandgap (4000 cm$^{-1}$, not shown in this figure) in VO$_2$, and serves as an indication of the Drude tail. We perform nano-IR broadband spectroscopy from 500-750cm$^{-1}$ to address the phonon responses (red bar). All the phonon peaks are below 750 cm-1, the highest two can be measured by broadband nano-IR spectroscopy, as discussed in the main text. Therefore in the blue bar region we effectively image the electronic response (insulator to metal transition) while in the red bar region we effectively measure the phonon response (monoclinic to rutile phase transition). We also confirm our imaging experiments at 1500 cm$^{-1}$ (not shown) and the conclusions remain the same.

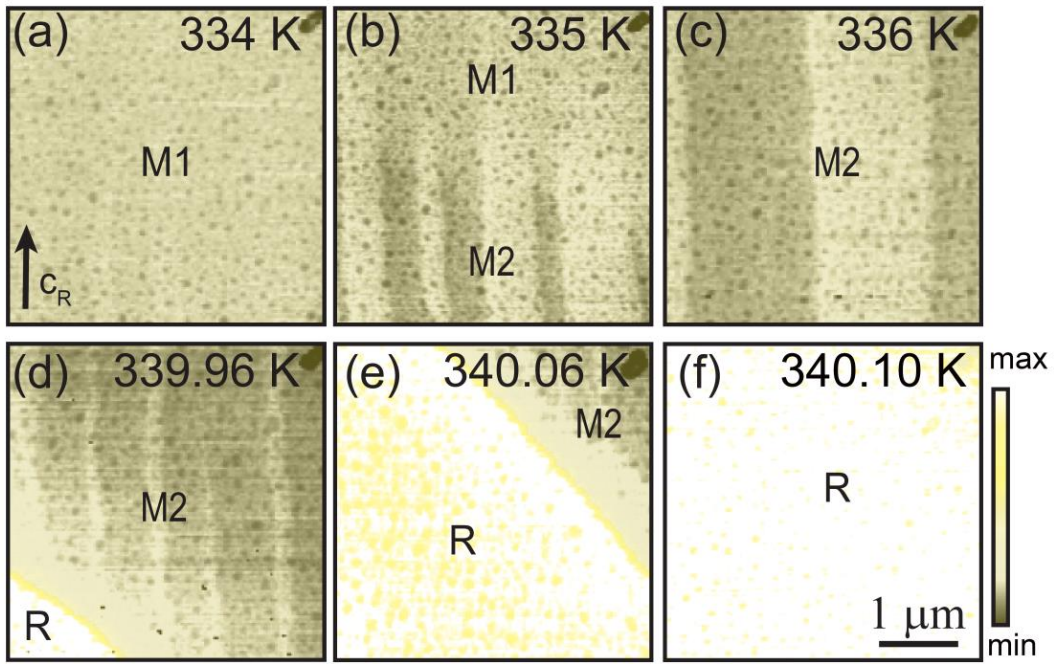

**Figure S2** IR scattering amplitude $S_3$ of $VO_2$ bulk single crystals acquired at different temperatures with enhanced contrast for insulating states (M1 and M2).